\documentclass[%
reprint,
superscriptaddress,
 amsmath,amssymb,
 aps,
 prb,
]{revtex4-2}
\usepackage{graphicx}% Include figure files
\usepackage{dcolumn}% Align table columns on decimal point
\usepackage{bm}% bold math
\usepackage{siunitx}
\usepackage{multirow}
\usepackage{tikz}
\usepackage{hyperref}% add hypertext capabilities
\usepackage{caption}
\usepackage{subfig}      % For creating subfloats
\usepackage{threeparttable}
\usepackage{booktabs} % For prettier tables
\hypersetup{colorlinks=true,citecolor=blue,linkcolor=red,urlcolor=blue}
\usepackage[mathlines]{lineno}% Enable numbering of text and display math
\usepackage{float}
\usepackage[english]{babel}
\newcommand{\NaKtwo}{NaK\textsubscript{2}Sb}
\newcommand{\Natwo}{Na\textsubscript{2}KSb}
\newcommand{\cNaKtwo}{cNaK\textsubscript{2}Sb}
\newcommand{\cNatwo}{cNa\textsubscript{2}KSb} 
\newcommand{\hNaKtwo}{hNaK\textsubscript{2}Sb}
\newcommand{\hNatwo}{hNa\textsubscript{2}KSb}

\DeclareSIUnit{\angstrom}{\textup{\AA}}
 \DeclareUnicodeCharacter{0308}{}
 
\begin{document}
%\title{Computationally predicted XANES of the $\text{Na}_{3-x}\text{K}_{x}\text{Sb}$ system}
\title{Ab initio X-ray Near-Edge Spectroscopy of Sodium-Based Multi-Alkali Antimonides}

% \author{Chung Xu$^1$, Richard Schier$^1$, and Caterina Cocchi$^{1,2,^*}$}
% \address{$^1$ Carl von Ossietzky Universit\"at Oldenburg, Institute of Physics, 26129 Oldenburg, Germany}
% \address{$^2$ Carl von Ossietzky Universit\"at Oldenburg, Center for Nanoscale Dynamics (CeNaD), 26129 Oldenburg, Germany}
% \address{$^*$ Author to whom any correspondence should be addressed}
% % \ead{caterina.cocchi@uni-oldenburg.de}
\author{Chung Xu}
 % \altaffiliation[Also at ]{Physics Department, XYZ University.}%Lines break automatically or can be forced with \\
 \email{chung.ping.xu@uni-oldenburg.de}
\affiliation{%
 Carl von Ossietzky Universit\"at Oldenburg, Institute of Physics, 26129 Oldenburg,
Germany
}%
\affiliation{%
 Department of Physics, University of Warwick, Coventry, CV4 7AL, United Kingdom
}%

\author{Richard Schier}
 \email{richard.schier@uni-oldenburg.de}
\affiliation{%
 Carl von Ossietzky Universit\"at Oldenburg, Institute of Physics, 26129 Oldenburg,
Germany
}%
\affiliation{Friedrich-Schiller Universit\"at Jena, Institute for Condensed Matter Theory and Optics, 07743 Jena, Germany}

\author{Caterina Cocchi}
\email{caterina.cocchi@uni-jena.de}
\affiliation{%
 Carl von Ossietzky Universit\"at Oldenburg, Institute of Physics, 26129 Oldenburg,
Germany
}%
\affiliation{Friedrich-Schiller Universit\"at Jena, Institute for Condensed Matter Theory and Optics, 07743 Jena, Germany}

\date{\today}
%\keywords{multi-alkali antimonides, sodium-potassium antimonides, semiconductors, photocathodes, photovoltaics, density-functional theory, many-body perturbation theory, electron sources, photoelectric effect, Bethe-Salpeter Equation, XANES}%Use showkeys class option if keyword

\begin{abstract}
Multi-alkali antimonides (MAAs) are promising materials for vacuum electron sources.  While sodium-based MAAs have demonstrated superior characteristics for ultrabright electron sources, their synthesis remains challenging, often resulting in mixed stoichiometries and polycrystalline domains. To address this complexity and guide the characterization of experimentally grown photocathodes, we present a comprehensive theoretical study of the X-ray near-edge spectroscopy (XANES) of four ternary MAAs: cubic \Natwo~and hexagonal \NaKtwo, representing the experimentally known phase of each stoichiometry, as well as hexagonal \Natwo~and cubic \NaKtwo, two computationally predicted polymorphs. Employing state-of-the-art \textit{ab initio} methods based on all-electron density-functional theory and the solution of the Bethe-Salpeter equation (BSE), we compute and analyze the XANES spectra at the sodium and potassium K-edges, potassium L$_{2,3}$-edge, and antimony K and L$_2$-edges. Our analysis reveals distinct spectral fingerprints for the experimentally known phases, cubic \Natwo~and hexagonal \NaKtwo, particularly at the sodium K-edge and potassium L$_{2,3}$-edge, providing useful indications for their identification in complex samples. By comparing BSE spectra against their counterparts obtained in the independent-particle approximation, we address the role of excitonic effects, highlighting their influence on the near-edge features, especially for excitations from the alkali metals. Our findings offer a useful theoretical benchmark for the characterization and diagnostics of sodium-based MAA photocathodes, complementing experiments on resolved phases and providing the spectral fingerprints of computationally predicted phases that could emerge in polycrystalline samples. 
\end{abstract}

\maketitle

\section{\label{sec:intro}Introduction}
Multi-alkali antimonides (MAAs) are an established class of semiconducting materials for vacuum electron sources~\cite{dowe+10nimpra,ADVANCES_ESOURCE+MUSUMECI2018209,PhysRevAccelBeams.21.113401}. Thanks to their favorable properties, including a narrow band gap, low electron affinity, and sensitivity to visible light, cesium antimonide photocathodes have been investigated for decades~\cite{jack-wach57prsla,NAKSBPHASES_MCCARROLL196030,mich+94nimpra,dibo+97nimpra,ding+17prab}. After the first seminal studies of the past century~\cite{sommer-1955,spic58pr,CsKSb_nathan-1967,ghosh-1978}, these materials are still the subject of intensive experimental~\cite{mamu+17prab,ding+17jap,dai+20electronics,panu+21nimpra,pavl+22apl,kach+23apl,guo+25sr} and computational research~\cite{guo14mre,Cs_pressure_KALARASSE20101732,xie+16prab,gupt+17jap,Cocchi_2019,Cs_cocchi-2019,cocc20pssrrl,anto+20prb,wu-gano23jmca,csk2sb_PBE,Santana-Andreo_2024}. More recently, Na-based MAAs have demonstrated superior characteristics for ultrabright electron sources, such as near-infrared optical response \cite{moha+23micromachines} and enhanced thermal emittance compared to other MAAs~\cite{EBEAM_10.1063/1.4945091,BIALKALI_MOTTA}. The experimental studies exploring these materials have been complemented by \textit{ab initio} investigations~\cite{yala+18jmmm,khan+21ijer, amador, yue+22prb,schi+24ats, xucomppredict}, confirming their near-infrared absorption threshold and overall favorable characteristics as photocathodes.

Despite these encouraging results, the growth of Na-based MAA photocathodes remains challenging~\cite{Na2KSb_QE_experiment,dube+25arxiv}, and samples often exhibit coexisting stoichiometries and polycrystalline domains~\cite{Na2KSb_QE_experiment,SPIN_SOURCE_PhysRevLett.129.166802}. 
Based on this complexity, non-invasive diagnostic techniques are of primary importance to characterize the samples and identify specific phases and compositions. X-ray spectroscopy is particularly suited for this purpose thanks to its sensitivity to the local atomic environment~\cite{bran+00jssc,cocc+16prb,vorw+18jpcl,frat+18materials,wu+21prb,olov-magn22jpcc,wibo+23jpcc,zimi+24pccp}.
Recent experimental advances~\cite{grec+23nrmp,grec+25nrm} have been accompanied by the refinement of \textit{ab initio} methods that are capable of describing the excitation process, explicitly taking into account electron-hole interactions~\cite{vorw+17prb,Vorwerk_2019,benf+21cpc,besl21wircms,vorw+22pccp}. Solving the Bethe-Salpeter equation (BSE) on top of all-electron density-functional theory (DFT)~\cite{olov+09prb,vorw+17prb,Vorwerk_2019} is currently considered the state-of-the-art approach to compute X-ray absorption spectra from first principles. Corresponding results provide useful indications to identify the spectral fingerprints of complex materials, not only by complementing experiments on the resolved phases but also by predicting the features of computationally discovered polymorphs that could appear as metastable phases in polycrystalline samples.

In this work, we employ DFT and BSE to investigate X-ray near-edge spectra (XANES) of four ternary MAAs with chemical compositions \Natwo~and \NaKtwo~with either hexagonal or cubic lattice. Cubic \Natwo~and hexagonal \NaKtwo~represent the experimentally known phases of the respective crystals~\cite{NAKSBPHASES_MCCARROLL196030} while hexagonal \Natwo~and cubic \NaKtwo~are computationally predicted polymorphs that have been characterized computationally in recent work~\cite{xucomppredict}.
We compute the XANES of all four materials, exploring excitations from different core levels experimentally accessible by hard or soft X-rays.
We compare the spectral features of the different compounds, highlighting the fingerprints that enable distinguishing one phase and/or stoichiometry from the other. We additionally discuss the excitonic character of the absorption resonances through the comparison between BSE spectra and their counterparts computed in the independent-particle approximation (IPA), where electron-hole Coulomb interactions are neglected. This assessment offers insight into the characteristics of the core-level excitations and provides additional information on the electronic structure of these crystals. Our results offer a valuable tool for the diagnostics and characterization of Na-based MAA photocathodes.

%%%%%%%%%%%%%%%%%%%%%%%%%%%%%%%%%%%%%%%%%%%%%%%%%%%%%%%
\section{Methods}
\subsection{\label{sec:theory}Theoretical Background}%
The results presented in this work are obtained in the framework of all-electron DFT~\cite{exciting_Gulans_2014} and many-body perturbation theory~\cite{Vorwerk_2019}. After solving the Kohn-Sham equations~\cite{KS_PhysRev.140.A1133}, we compute the XANES from the BSE~\cite{vorw+17prb,Vorwerk_2019} mapped into the eigenvalue problem
\begin{equation}\label{eqn:BSE}
\sum_{o^{\prime} u^{\prime} \mathbf{k}^{\prime}} \hat{H}_{ou \mathbf{k}, o^{\prime} u^{\prime} \mathbf{k}^{\prime}}^{\mathrm{BSE}} A_{o^{\prime} u^{\prime} \mathbf{k}^{\prime}}^\lambda=E^\lambda A_{ou \mathbf{k}}^\lambda,
\end{equation}
where the eigenvalues $E^\lambda$ are the excitation energies, and the eigenvectors $A^\lambda$ contain information about the character and composition of the excitations. They both enter the imaginary part of the macroscopic dielectric tensor, 
\begin{equation}\label{eqn:macrdielectric}
\Im \varepsilon_M=\frac{8 \pi^2}{\Omega} \sum_\lambda\left|\mathbf{t}^\lambda\right|^2 \delta\left(\omega-E^\lambda\right),
\end{equation}
where $\Omega$ is the unit cell volume, $\omega$ is the angular frequency of the incoming photon, and the $\mathbf{t}^\lambda$ are the transition coefficients between initially occupied core levels ($o$) and unoccupied conduction states ($u$),
\begin{equation}\label{eqn:transitioncoeffs}
\mathbf{t}^\lambda=\sum_{ou\mathbf{k}} A_{ou\mathbf{k}}^\lambda \frac{\langle o\mathbf{k}|\hat{\mathbf{p}}| u \mathbf{k}\rangle}{\epsilon_{u \mathbf{k}}-\epsilon_{o}},
\end{equation}
with corresponding KS energies appearing in the denominator.

The BSE Hamiltonian in Eq.~\eqref{eqn:BSE} is
\begin{equation}
    \hat{H}^\text{BSE} = \hat{H}^{\rm diag}+2\hat{H} ^{\rm x} +\hat{H}^{\rm dir},
    \label{eq:H_BSE}
\end{equation}
where $\hat{H}^\text{diag}$ accounts for vertical transitions between core and conduction states, 2$\hat{H}^{\rm x}$ represents the repulsive exchange interaction between photoexcited electron and hole (the multiplication factor 2 arises from spin-degeneracy), and $\hat{H}^{\rm dir}$ embeds the statically screened electron-hole Coulomb potential (direct term). In its full form, $\hat{H}^\text{BSE}$ includes the key physical interactions that are crucial for an accurate description of XANES spectra. However, to reveal and quantify excitonic effects and associate the spectral features to the (unoccupied) electronic structure of the material described by the projected density of states (PDOS), it is instructive to compare the BSE spectra with their counterparts computed in the IPA, obtained by diagonalizing $\hat{H}^\text{IPA} \equiv \hat{H}^{\rm diag}$. BSE and IPA spectra are usually different unless electron-hole correlations are negligible.

\subsection{\label{sec:param}Computational Details}%
All calculations are performed with \texttt{exciting}, a full-potential all-electron code implementing the linearized augmented plane waves and local orbitals method for DFT and BSE~\cite{exciting_Gulans_2014,Vorwerk_2019}. The explicit treatment of core electrons makes this software ideally suited to address the central problem of this work~\cite{vorw+17prb}.
To ensure convergence, we set the muffin-tin radius of both Na and K atoms to 2.0~bohr and that of Sb to 2.2~bohr. The plane-wave cutoff for the basis set, defined in this framework as the product between the smallest muffin-tin radius among the considered atomic species and the maximum length for the \textbf{k}-vector~\cite{exciting_Gulans_2014}, is fixed to 8.0 and 8.5 for the cubic  and hexagonal crystals, respectively.
The exchange-correlation potential of DFT is approximated using the spin-unpolarized Perdew-Burke-Ernzerhof functional for solids (PBEsol)~\cite{GGA_PBE_SOL}. In the DFT runs, the Brillouin zones of the cubic and hexagonal crystals are sampled with $10\times 10\times 10$ and $8 \times 8 \times 4$ $\mathbf{k}$-meshes, respectively. 

The BSE is constructed and solved in the Tamm-Dancoff approximation~\cite{Vorwerk_2019}, computing the screened Coulomb interaction in the random phase approximation with 100 empty states. An energy cutoff of 1.5~Ha is taken for the local fields. The BSE is solved with 10 (12) empty states for the cubic (hexagonal) crystals, adopting a $\Gamma$-shifted \textbf{k}-mesh with $8 \times 8 \times 8$ points for the cubic crystals and $8 \times 8 \times 4$ for the hexagonal ones to obtain a proper resolution of the spectral features.
 A Lorentzian broadening of \SI{100}{\milli\eV} is used to visualize all spectra. For the cubic crystals, we plot the unique non-zero term of Eq.~\eqref{eqn:macrdielectric}, while for the hexagonal crystals, we take the average of the three Cartesian components.

\begin{figure}[h]
    \centering
    \includegraphics[width=0.48\textwidth]{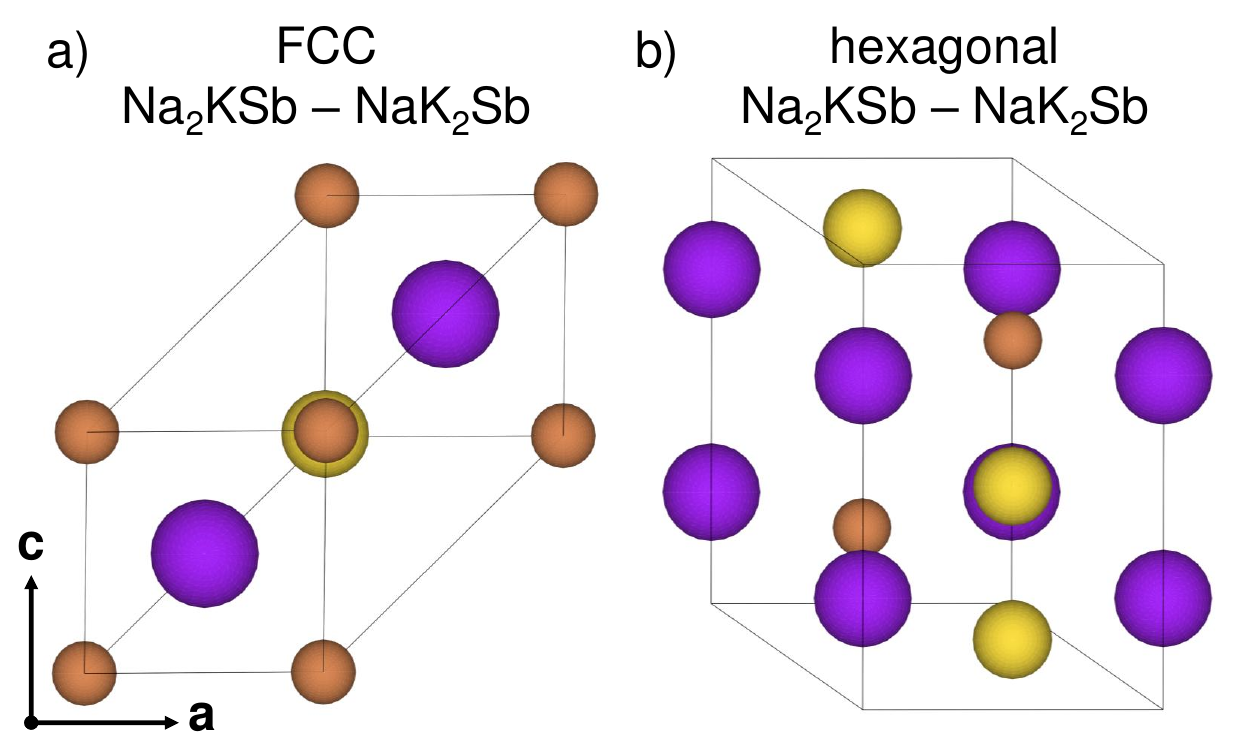}
    \caption{a) Face-centered cubic (FCC) and b) hexagonal unit cells of the Na$_2$KSb and NaK$_2$Sb crystals considered in this work. Sb atoms are in bronze while the alkali species are in yellow and blue. Graphs produced with the visualization software VESTA~\cite{Momma:db5098}.}
    \label{fig:structure}
\end{figure}

The bulk crystals investigated in this work are taken from the Open Quantum Materials Database (OQMD)~\cite{OQMD}, structure numbers: 7813090 for hexagonal \Natwo~(\hNatwo), 42945 for cubic \Natwo~(\cNatwo), 1245926 for cubic \NaKtwo~(\cNaKtwo), and 7751654 for hexagonal \NaKtwo~(\hNaKtwo). The cubic phases of both Na$_2$KSb and NaK$_2$Sb are face-centred cubic Bravais lattices with Sb atoms at Wyckoff position $(0,0,0)$ and alkali species at $(1/2,1/2,1/2)$ and $\pm(1/4, 1/4, 1/4)$, see Fig.~\ref{fig:structure}a. In the hexagonal crystals, the Sb atoms occupy the $(2/3,1/3,3/4)$ and $(1/3,2/3,1/4)$ Wyckoff positions while alkali atoms are at $(2/3,1/3,1/4 \pm 1/6)$, $(1/3,2/3,3/4 \pm 1/6)$, and $(0, 0, \pm 1/4)$, see Fig.~\ref{fig:structure}b. The initial structures taken from OQMD are subsequently relaxed with \texttt{exciting}, using the Broyden-Fletcher-Goldfarb-Shanno algorithm~\cite{head-zern85cpl} to minimize interatomic forces and a volume optimization based on the Birch-Murnaghan fit~\cite{birc+47pr,murn44pnas}. The same procedure was adopted in a previous related study~\cite{xucomppredict}, where further details can be found. 

%%%%%%%%%%%%%%%%%%%%%%%%%%%%%%%%%%%%%%%
\section{\label{sec:results}Results}
In the following, we analyze XANES spectra computed for the four MAA crystals considered in this work. 
Despite the importance of surface characteristics in the performance of photocathodes, we perform our analysis on bulk materials. This choice is motivated by two main reasons: first, XANES is inherently not a surface-sensitive technique; second, experimental samples are typically orders of magnitude thicker~\cite{mamu+15aplm} than the surface slabs accessible from first principles~\cite{wang+22ssc,schi+22prm}.

In Sec.~\ref{sec:Na-K}, we examine excitations from the sodium K-edge (1$s$ core electrons), targeting unoccupied electrons with $p$-orbital character. In Sec.~\ref{sec:K-K}, we inspect spectra computed from the potassium K-edge, whereby the 1$s$ electrons of this element are excited to unoccupied $p$-states. In Sec.~\ref{sec:K-L23}, we discuss excitations from the potassium L$_{2,3}$-edge, corresponding to transitions from the 2$p$ core electrons of this element to unoccupied levels bearing $s$ and $d$ character. In Sec.~\ref{sec:Sb-K}, we investigate the XANES spectra computed from the antimony K-edge, \textit{i.e.}, excitations from Sb 1$s$ electrons to unoccupied bands with Sb $p$ character. Finally, in Sec.~\ref{sec:Sb-L2}, we analyze the spectra obtained by exciting antimony 2$p$ core electrons to unoccupied states with Sb $s$ and $d$ character. Due to the large spin-orbit splitting (several tens of eV) in the Sb 2$p$ shell, we consider only the 2$p_{1/2}$ component of the XANES (L$_2$-edge), which yields equivalent signatures, albeit with different oscillator strength, as expected, than the L$_3$-edge~\cite{thol-vand88prb}.

%%%%%%%%%%%%%%%%%%%%%%%%%%%%%%%%%%%%%%%%%%
\subsection{Sodium K-edge}\label{sec:Na-K}
\begin{figure*}
    \centering
    \includegraphics[width=1\linewidth]{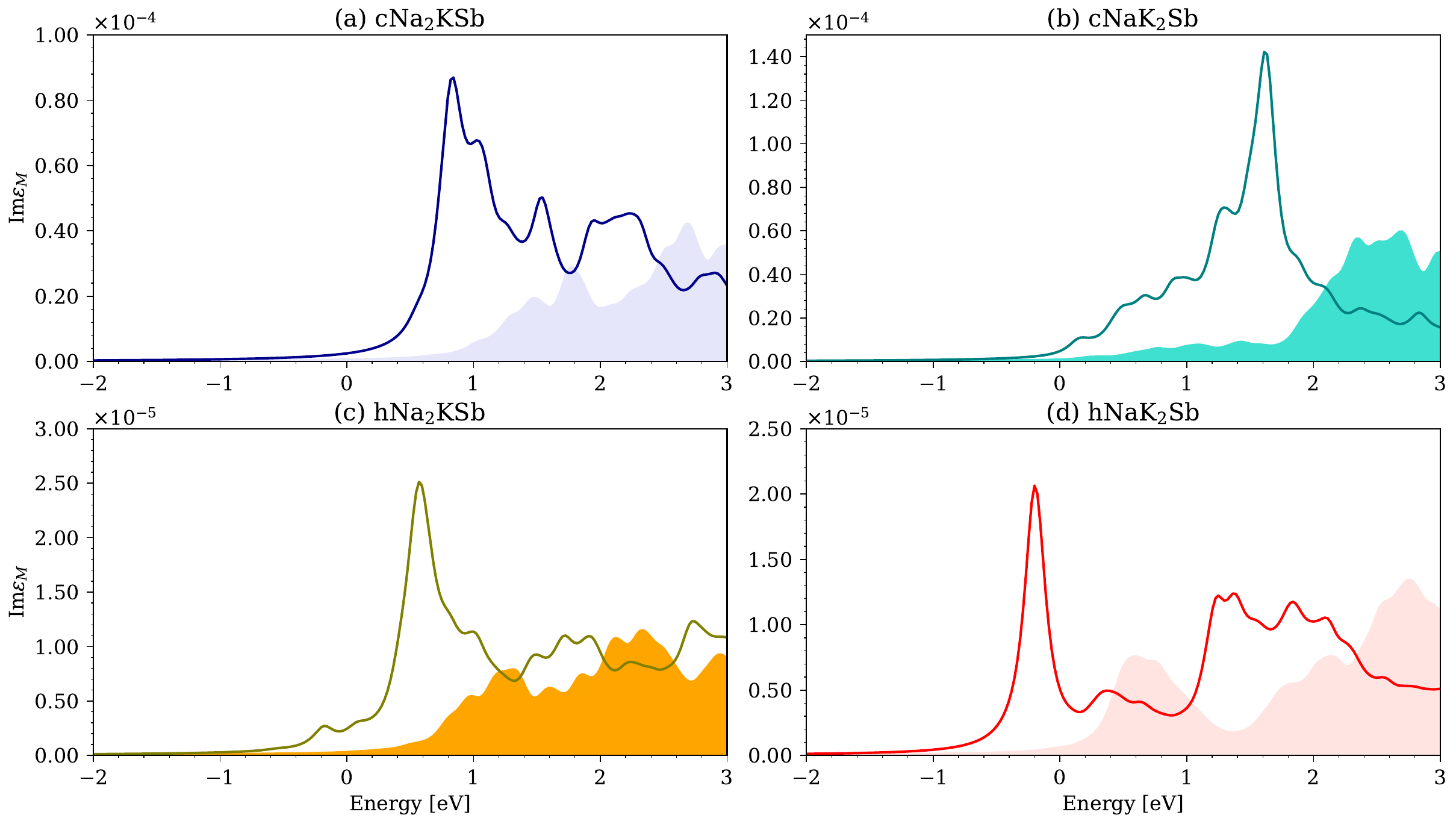}
    \caption{XANES of (a) cubic \Natwo, (b) cubic \NaKtwo, (c) hexagonal \Natwo, and (d) hexagonal \NaKtwo~computed from the sodium K-edge. Solid lines (shaded areas) represent BSE (IPA) spectra. }
    \label{fig:nak}
\end{figure*}

We begin our analysis with the inspection of the sodium K-edge spectra (Fig.~\ref{fig:nak}), which reveal notable similarities between the two \Natwo~phases (Fig.~\ref{fig:nak}a,c). Both spectra exhibit a high-intensity peak at low energies, with the maximum at 0.9~eV and 0.6~eV above the IPA onset for \cNatwo~and \hNatwo, respectively. The excitonic nature of this peak is evident from the comparison between the BSE and the IPA spectra. The former exhibits a strong peak at low energies (up to 1~eV from the IPA onset), where uncorrelated vertical transitions lead to extremely weak oscillator strength. This oscillator strength enhancement and a red shift of almost 1~eV are characteristic signatures of strong electron-hole correlations. In the spectrum of \hNatwo, a weak but distinct feature is identified around -0.1~eV and associated with a double-degenerate bound exciton with binding energy of 70~meV (Table~S1). Two pairs of degenerate excitons appear below the IPA onset of \cNatwo, with binding energies of 229~meV and 59~meV, but their oscillator strength is so weak that they do not produce any visible maximum in Fig.~\ref{fig:nak}a.

The spectra of the \NaKtwo~crystals exhibit remarkably different signatures. The XANES of \cNaKtwo~is characterized by a weak but increasingly intense onset culminating in the pronounced resonance at approximately 1.5~eV (Fig.~\ref{fig:nak}b). On the other hand, the spectrum of \hNaKtwo~is dominated by a sharp excitonic resonance below the IPA onset, given by two degenerate excitons with a binding energy of approximately 130~meV (Table~S1).
Comparison with the respective IPA spectra confirms the excitonic character of all these features, whereby electron-hole interactions are responsible for spectral weight enhancement at low energies and for a redshift of the order of 1~eV.
In the XANES of \cNaKtwo, two pairs of degenerate excitons appear below the IPA onset, with binding energies of 574~meV and 55~meV (Table~S1). In analogy with the \Natwo~compounds, their oscillator strength is negligible due to the minimal contribution of the Na $p$-states to the bottom of the conduction band~\cite{xucomppredict}, giving rise to an energetically broad but weak band (see Fig.~S1b).

The similarities between the spectra of the materials with equal stoichiometry and different crystal structures are expected to facilitate the identification of specific compositions in polycrystalline samples. Likewise, the distinct spectral features of the experimentally resolved crystals, \cNatwo~and \hNaKtwo~(Fig.~\ref{fig:nak}a,d), will promote their recognition in mixed samples.

%%%%%%%%%%%%%%%%%%%%%%%%%%%%%%%%%%%%%%%%%%%%
\subsection{Potassium K-edge}\label{sec:K-K}

\begin{figure*}
    \centering
    \includegraphics[width=1\linewidth]{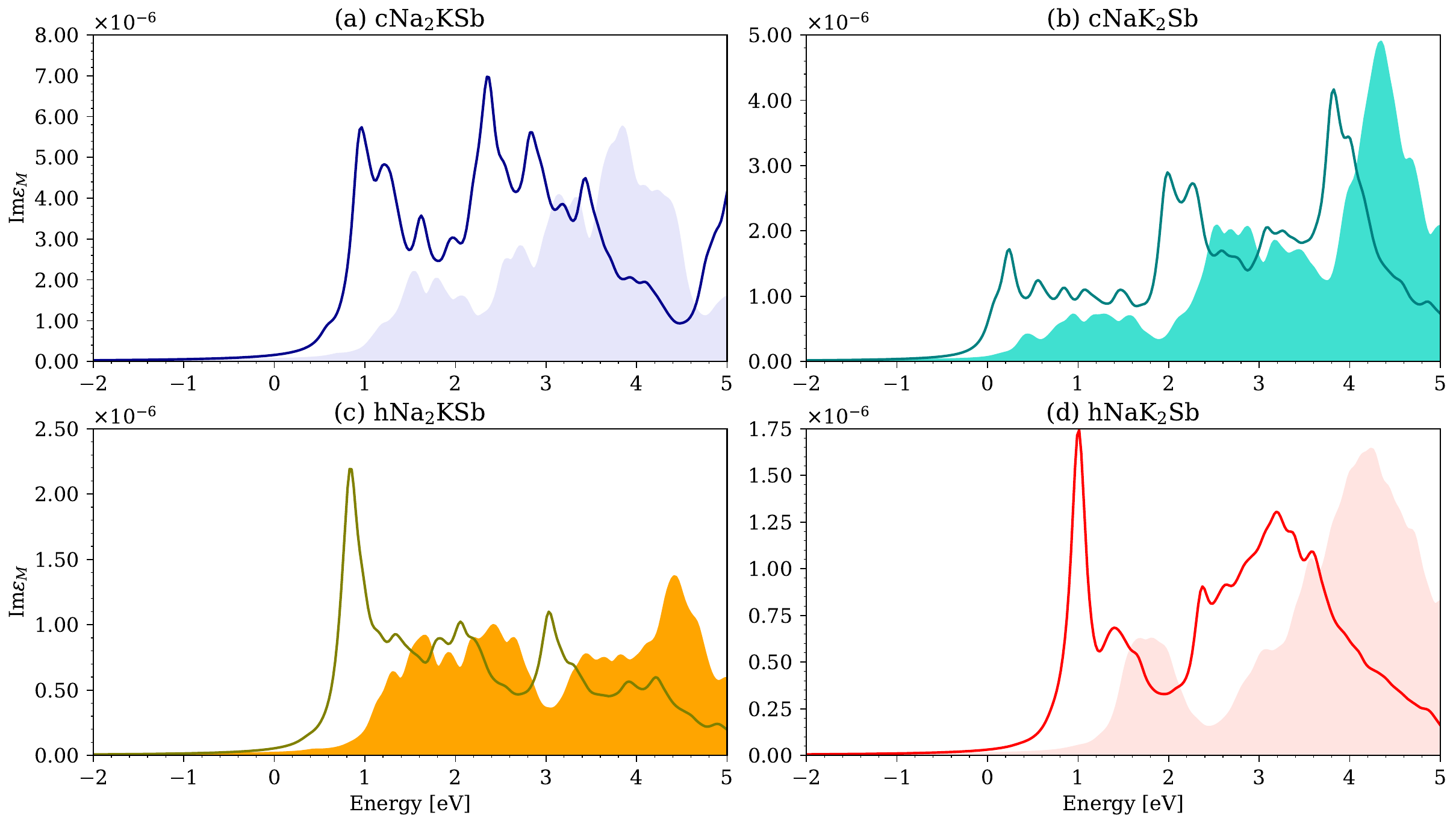}
    \caption{XANES of (a) cubic \Natwo, (b) cubic \NaKtwo, (c) hexagonal \Natwo, and (d) hexagonal \NaKtwo~computed from the potassium K-edge. Solid lines (shaded areas) represent BSE (IPA) spectra. }
    \label{fig:kk}
\end{figure*}

In analogy to the Na K-edge XANES, the results obtained for the cubic and hexagonal phases of \Natwo~are very similar, see Fig.~\ref{fig:kk}a,c. Both spectra are characterized by a broad absorption onset with several maxima covering the first 3~eV above the IPA onset, and two degenerate dark excitons below the IPA onset with binding energies of 111~meV in \cNatwo~and 60~meV in \hNatwo (Table~S1). Electron-hole correlations enhance the oscillator strength in this region, leading to a redistribution of the spectral weight to lower energies upon inclusion of electron-hole correlations~\cite{cocc+16prb}, as from the Na K-edge. The IPA spectra of the \Natwo~phases reproduce the characteristics of the K $p$-contributions to the PDOS, especially concerning the stronger maxima around 3 and 4~eV, see Fig.~S2a,c.

The XANES of the computationally predicted \cNaKtwo~crystal is characterized by broad excitations at low energies (Fig.~\ref{fig:kk}b). However, the oscillator strength of the first peak is weaker than in \cNatwo~and the BSE spectrum retains the overall spectral shape of its IPA counterpart, confirming the less prominent role of excitonic effects. On the other hand, the spectrum of \hNaKtwo~is dominated by a sharp excitonic peak at the onset (Fig.~\ref{fig:kk}d), stemming from the broad feature in the IPA spectrum between approximately 1.2 and 2.2~eV which finds a direct counterpart in the K $p$-contributions to the PDOS of this material (Fig.~S2d). The absorption dip appearing between this maximum in the IPA spectrum and the following stronger region of absorption stems directly from the K $p$-PDOS characteristics (see Fig.~S2d and Ref.~\cite{amador}). This minimum is also visible in the XANES computed from the BSE, where it is red-shifted by about 0.5~eV (Fig.~\ref{fig:kk}d). The comparison between the sharp excitonic resonance in the BSE result and the energetically distributed oscillator strength in the IPA spectrum allows assigning the former a binding energy of approximately 800~meV. The sharp excitonic maximum has a shoulder at higher energies and is followed by a broader region with non-negligible oscillator strength, red-shifted by about 1~eV from its IPA counterpart (Fig.~\ref{fig:kk}d). The presence of this prominent excitonic feature in the XANES spectra of \hNaKtwo~makes this experimentally resolved phase well detectable in polycrystalline samples.

%%%%%%%%%%%%%%%%%%%%%%%%%%%%%%%%%%%%%%%%%%%%%%%%%%%%%%%%
\subsection{Potassium L$_{2,3}$-edge}\label{sec:K-L23}

\begin{figure*}
    \centering
    \includegraphics[width=1\linewidth]{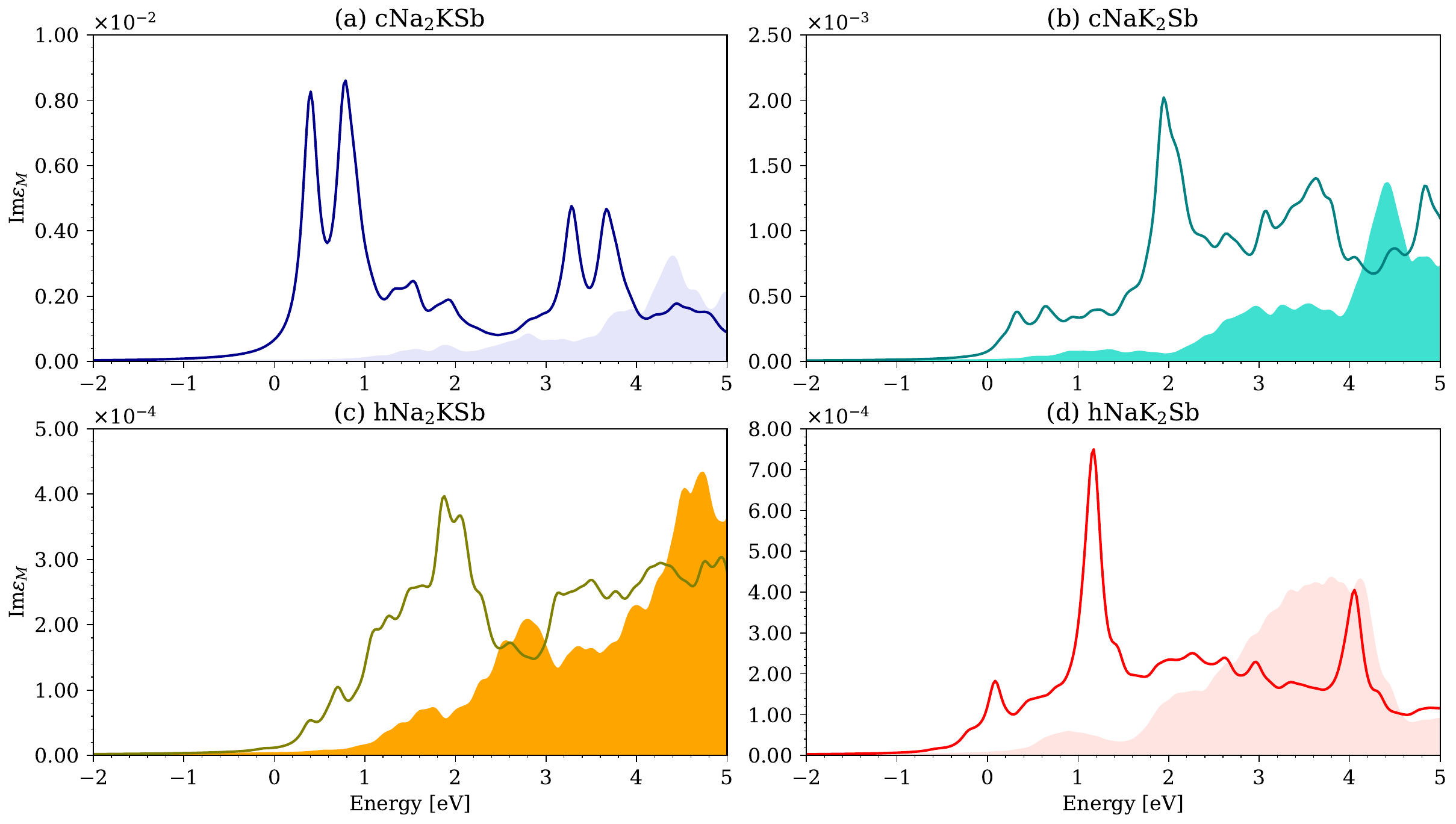}
    \caption{XANES of (a) cubic \Natwo, (b) cubic \NaKtwo, (c) hexagonal \Natwo, and (d) hexagonal \NaKtwo~computed from the potassium L$_{2,3}$-edge. Solid lines (shaded areas) represent BSE (IPA) spectra. }
    \label{fig:K_L23}
\end{figure*}

We continue our analysis with the XANES computed from the potassium L$_{2,3}$-edge. The result obtained for the experimentally known cubic phase of \Natwo~is dominated by two intense peaks at the onset (Fig.~\ref{fig:K_L23}a), which offer a valuable spectroscopic signature to identify this phase in a polycrystalline sample. The excitonic nature of these maxima is evident from the comparison with the IPA spectrum, which features weak oscillator strength distributed up to almost 3.5~eV (Fig.~\ref{fig:K_L23}a). The replicas of these peaks, appearing approximately 3~eV above the first ones, stem from transitions from 2$p_{1/2}$ electrons and, as expected, their oscillator strength is approximately $2/3$ of those generated by excitations from the 2$p_{3/2}$ core states. 

The XANES of the computationally predicted cubic \NaKtwo~is characterized by very weak peaks at the onset and by a strong maximum at 2~eV (Fig.~\ref{fig:K_L23}b). The replica of this feature is visible at about 5~eV, but it would be hardly detectable in experiments due to the relatively large oscillator strength of overlapping excitations between 3 and 4~eV. Excitonic effects are pronounced and manifest themselves through a strong red-shift of the spectral weight upon inclusion of electron-hole correlations. The IPA spectrum, in turn, assimilates the features of the K $d$-states in the PDOS, which dominate over the contributions from K $s$-electrons (Fig.~S3b).

The spectrum of the other computationally predicted compound, \hNatwo, is again characterized by excitations of increasing strength at the onset up to approximately 2~eV (Fig.~\ref{fig:K_L23}c). The spectral shape in the BSE result somehow reflects the IPA spectrum, suggesting that electron-hole interactions mainly red shift the excitations by about 1~eV and enhance their oscillator strength toward lower energies. In this case, the high-energy replica of the first excitations is not clearly visible, due to other bright excitations appearing in the same energy range. Similar to \cNaKtwo, the target states for the core-level transitions are mainly given by the K $d$-states, which dominate over the $s$-orbital contributions in the relevant energy region (Fig.~S3c).

The XANES computed for \hNaKtwo, the other experimentally known phase, is dominated by a sharp peak slightly about 1~eV, see Fig.~\ref{fig:K_L23}d. The excitonic nature of this feature is testified not only by its very strong oscillator strength but also by the absence of a direct counterpart in the IPA spectrum. A replica of this intense maximum is visible around 4~eV, \textit{i.e.}, about 3~eV above the first one, with an oscillator strength approximately $2/3$ of the lower-energy resonance, as expected. Interestingly, a weaker but equally sharp peak is visible a few tens of meV above the onset. Its higher-energy replica is not visible, again due to overlap with other bright states. In this case, K $s$-orbital contributions are more pronounced in the unoccupied region targeted by the relevant core-level transitions (Fig.~S3d), but, given the shape of the IPA spectrum, excitations to K $d$-states are still expected to be dominant.

All compounds are characterized by four degenerate excitations below the IPA onset. Their binding energies are on the order of 60~meV in both \cNaKtwo~and \hNatwo, while they are almost twice as large in the experimentally known phases \cNatwo~and \hNaKtwo~(Table~S1). The oscillator strength of these bound excitons is negligible in the spectra of the two cubic polymorphs, where no signatures can be identified in Fig.~\ref{fig:K_L23}a,c. On the other hand, in the XANES of the two hexagonal phases, we notice very weak maxima below the IPA onset, with the one in \hNatwo~being hardly visible (Fig.~\ref{fig:K_L23}b), while the one in \hNaKtwo~appearing as a shoulder of the peak centered around 0~eV (Fig.~\ref{fig:K_L23}d).

The peculiar spectral characteristics of \cNatwo~and \hNaKtwo~make these two experimentally resolved phases expectedly well detectable in polycrystalline samples. In particular, the striking differences between the two spectra in Fig.~\ref{fig:K_L23}a and Fig.~\ref{fig:K_L23}d, namely the presence of two low-energy sharp peaks in the former and a single intense low-energy maximum in the latter, is anticipated to promote the identification of either compound in a mixed sample.

%%%%%%%%%%%%%%%%%%%%%%%%%%%%%%%%
\subsection{Antimony K-edge}~\label{sec:Sb-K}
\begin{figure*}[ht]
    \centering
    \includegraphics[width=1\linewidth]{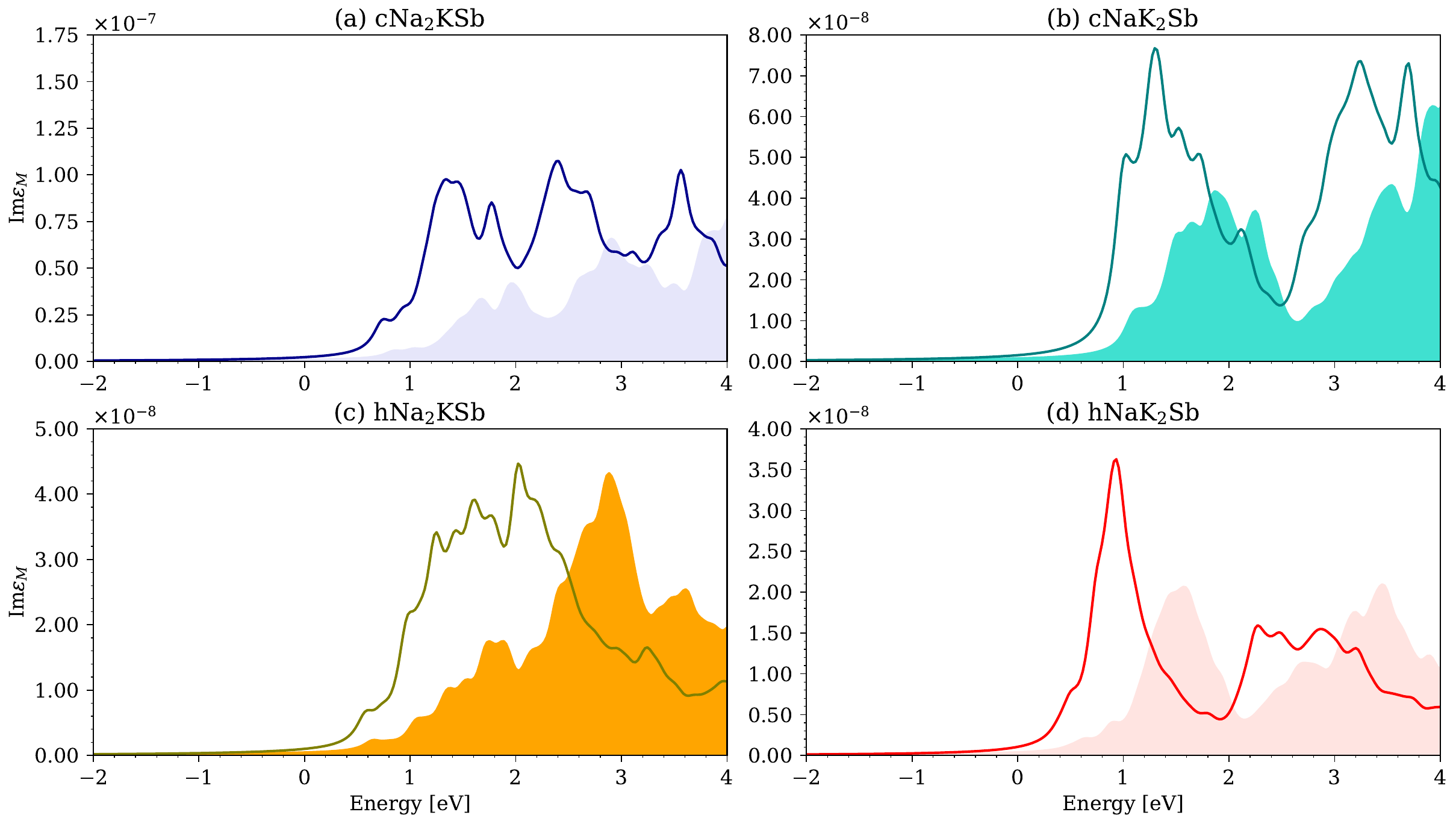}
    \caption{XANES of (a) cubic \Natwo, (b) cubic \NaKtwo, (c) hexagonal \Natwo, and (d) hexagonal \NaKtwo~computed from the antimony K-edge. Solid lines (shaded areas) represent BSE (IPA) spectra. }
    \label{fig:SbK}
\end{figure*}

The XANES calculated from the Sb K-edge exhibit pronounced similarities for each considered stoichiometry. The cubic and hexagonal phases of \Natwo~are characterized by relatively weak oscillator strength at the onset and larger spectral weight at higher energies, around 1~eV (Fig.~\ref{fig:SbK}a,c). In the BSE spectrum of \cNatwo~(Fig.~\ref{fig:SbK}a), the lowest-energy region is dominated by four relatively broad maxima mimicking the features of the IPA spectrum, which in turn reflects the Sb $p$-orbital contributions to the conduction region (Fig.~S4), including a dip between the two unoccupied Sb $p$-state manifolds~\cite{amador}. In the spectrum of the computationally predicted hexagonal phase of \Natwo, the oscillator strength at low energies is focused across a narrow range of approximately 1.5~eV (Fig.~\ref{fig:SbK}c). The first weak peak appears around 0.6~eV in both the BSE and IPA spectra, suggesting an almost negligible influence of electron-hole correlations. It is followed by a broad distribution of bright excitations with the most intense maximum at 2~eV. The spectral features characterizing the BSE result are almost identically reproduced in the IPA spectrum, only shifted by about 1~eV. This finding suggests that excitonic effects in this material manifest themselves in the above-mentioned sizeable red shift of excitation energies but do not alter the spectral shape. In both \Natwo~polymorphs, two degenerate excitations are found below the IPA onset. Both of them are dark, being two orders of magnitude weaker than the strongest resonances. Their binding energy is 105~meV in \cNatwo~and 59~meV in \hNatwo, see Table~S1. The systematically lower intensity of excitations from the Sb K-edge compared to those analyzed above (\textit{e.g.}, from the K L$_{2,3}$-edge) is ascribed to the negligible contribution of Sb $p$ states to the bottom of the conduction region of both materials (Fig.~S4a,c)~\cite{amador,xucomppredict}. 

The XANES of both \NaKtwo~phases are characterized by similar spectral fingerprints, which are in turn quite different from those of \Natwo. In the computationally predicted cubic polymorph, the absorption from the Sb K-edge is dominated by two broad regions ($>$1~eV each) of intense excitations, see Fig.~\ref{fig:SbK}b, both red-shifted by about 200~meV compared to the IPA results, which mirrors the salient features of the Sb $p$-orbital contributions to the PDOS (Fig.~S4b), including the small gap between the first and second absorption region around 2.8~eV in Fig.~\ref{fig:SbK}b~\cite{amador}. While the first manifold almost doubles its oscillator strength upon inclusion of electron-hole correlations, this effect is less pronounced for the higher-energy maxima. The spectrum of the experimentally resolved hexagonal structure of \NaKtwo~is dominated by an intense peak at about 1~eV (Fig.~\ref{fig:SbK}d). This strong excitation is preceded by a weak shoulder at approximately 0.5~eV. Higher-energy excitations have about half the oscillator strength of the aforementioned intense peak and form a broad band spanning almost 2~eV. The peaks in the BSE spectrum mirror those computed from the IPA -- reflecting the contributions of the Sb $p$-states to the PDOS (see Fig.~S4d and Ref.~\cite{amador}) -- and the red shift associated with their binding energy is on the order of 800~meV. However, only the intensity of the first resonance is significantly affected by the inclusion of electron-hole correlations. 
The sharp resonance in the spectrum of \hNaKtwo~represents a relevant signature to identify this structure in a polycrystalline sample, although the large broadening expected for excitations from such a deep edge ($>$30~keV~\cite{desl+03rmp}) likely overshadows these features during measurements.

%%%%%%%%%%%%%%%%%%%%%%%%%%%%%%%%%%%%%%%%%%
\subsection{Antimony L$_{2}$-edge}\label{sec:Sb-L2}

\begin{figure*}
    \centering
        \includegraphics[width=1\linewidth]{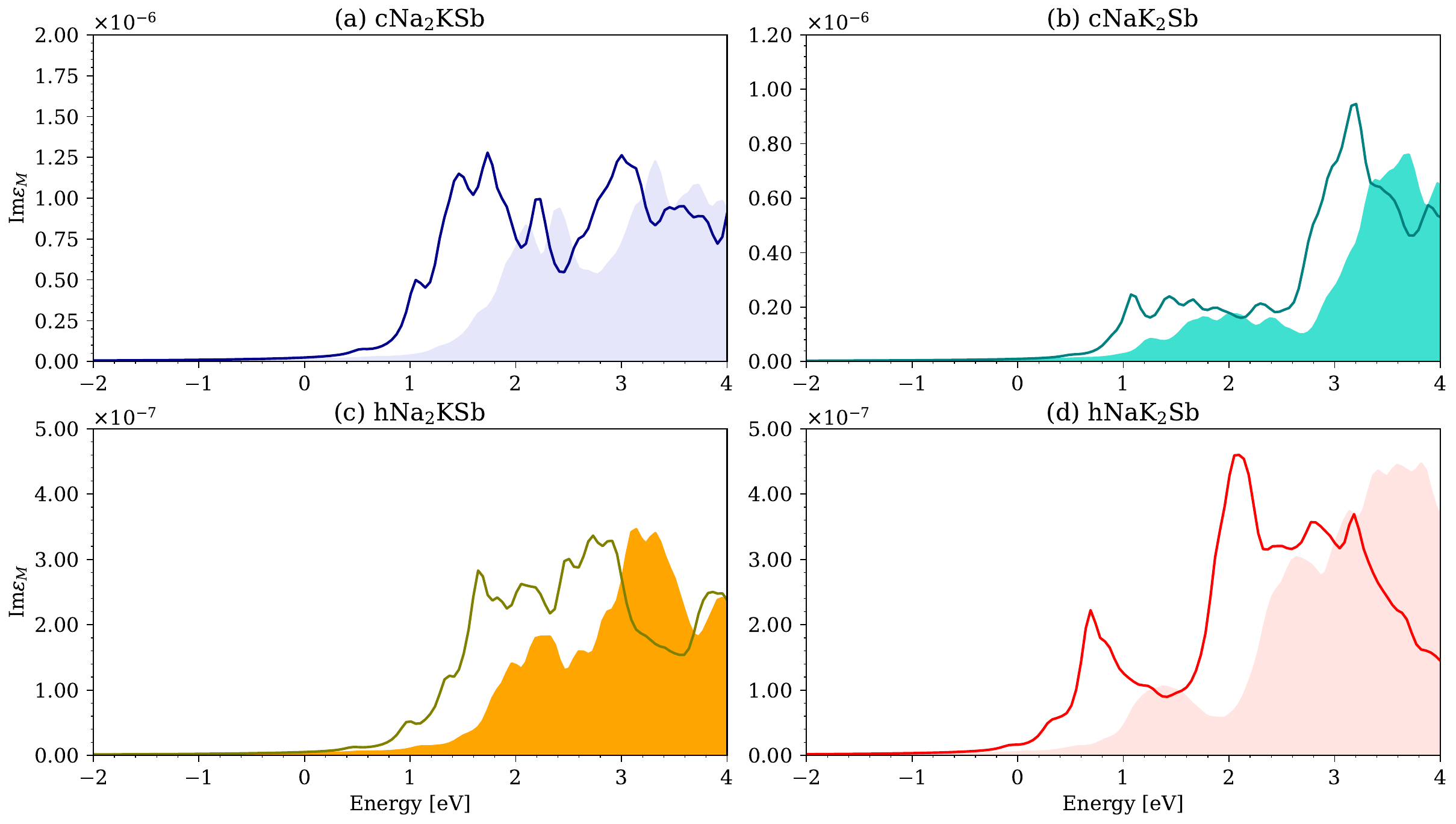}
    \caption{XANES of (a) cubic \Natwo, (b) cubic \NaKtwo, (c) hexagonal \Natwo, and (d) hexagonal \NaKtwo~computed from the antimony L$_2$-edge. Solid lines (shaded areas) represent BSE (IPA) spectra. }
    \label{fig:SbL2}
\end{figure*}

We conclude our analysis by inspecting the XANES computed from the Sb L$_2$-edge. Target states for the excited Sb 2$p$ electrons are unoccupied bands with Sb $s$ and $d$ character. As extensively discussed in Ref.~\citenum{amador} for the experimentally resolved phases \cNatwo~and \hNaKtwo, and in Ref.~\citenum{xucomppredict} for the computationally predicted polymorphs \hNatwo~and \cNaKtwo, Sb $s$-electrons dominate to the lowest unoccupied region of all materials while Sb $d$-states appear at slightly higher energies (see also Fig.~S5). The spectra of the two \Natwo~polymorphs are quite similar, see Fig.~\ref{fig:SbL2}a,c. They are both characterized by weak transitions at the onset and by stronger peaks above 1~eV. The spectral features from the BSE are present already in the IPA spectra, only with weaker oscillator strength (about $2/3$ for the first manifold of excitations up to approximately 2.2~eV) and an energy shift of about 700~meV in \cNatwo~and 400~meV in \hNatwo. The relatively large oscillator strength in the higher energy part of the IPA spectra is due to the combined contributions to transitions targeting both Sb $s$- and $d$-states appearing with similar weight a few eV above the conduction band minimum (Fig.~S5a,c).

The \NaKtwo~spectra display weak excitonic effects, with red shifts of the order of 200~meV in \cNaKtwo~(Fig.~\ref{fig:SbL2}b) and 600~meV in \hNaKtwo~(Fig.~\ref{fig:SbL2}d), and an oscillator strength of the first excitations only marginally amplified by electron-hole correlations. While the XANES of \cNaKtwo~hosts very weak excitations up to about 3~eV, the one of \hNaKtwo~exhibits a weak but distinguishable absorption band between 0.3 and 1~eV, and a stronger one between 2 and 3~eV. It is hard to expect that these features will enable the identification of specific stoichiometries and/or crystal structures in mixed samples. Similar to \Natwo, the larger oscillator strength in the IPA spectra at higher energies is due to the combined contributions of transitions targeting both Sb $s$- and $d$-states coexisting in that range of the conduction region (Fig.~S5b,d).

All compounds exhibit a doubly degenerate excitation below the IPA onset. Its oscillator strength is at least one order of magnitude lower than the transitions giving rise to the visible peaks in the XANES, while its binding energy is of the order of 60~meV in the two computationally predicted compounds (\cNaKtwo~and \hNatwo) and 120~meV in the experimental phases (\cNatwo~and \hNaKtwo), see Table~S1. Only in the spectrum of \hNaKtwo, the bound excitons give rise to a visible signal in the XANES (Fig.~\ref{fig:SbL2}d). Overall, the relatively weak spectral weight of these absorption spectra and the absence of any distinct resonance make it hard to expect the signatures from XANES to provide an effective tool to identify different stoichiometries or crystal phases from the Sb L$_{2,3}$-spectra of polycrystalline sodium potassium antimonide samples.

\section{\label{sec:conclusion}Summary, Discussion, and Conclusions}%
In summary, we investigated from first-principles many-body theory the XANES spectra of four Na-based ternary alkali antimonide phases, considering excitations from sodium, potassium, and antimony K-edge, potassium L$_{2,3}$-edge, and antimony L$_2$-edge. In this analysis, we included both the experimentally resolved cubic polymorph of \Natwo~and hexagonal phase of \NaKtwo~\cite{NAKSBPHASES_MCCARROLL196030,amador}, as well as the computationally predicted cubic \NaKtwo~and hexagonal \Natwo~\cite{xucomppredict}. By comparing the results obtained for all compounds, we identified the spectral fingerprints that can enable their identification in polycrystalline samples embedding different stoichiometries. From a fundamental viewpoint, we addressed the role of electron-hole correlations to gain further insight into the electronic structure of each material. 

The intense excitonic resonances dominating the onset of the sodium and potassium K-edge XANES of \hNaKtwo~represent a valuable tool to identify this crystal in mixed samples. Likewise, the potassium L$_{2,3}$-edge spectra of the two experimental phases, \cNatwo~and \hNaKtwo, are dominated by clear spectral signatures promoting their recognition in polycrystalline samples. In particular, the spectrum of \cNatwo~hosts two resonances at the onset with strong intensity and almost equal oscillator strength, while the XANES of \hNaKtwo~features a distinct strong peak about 1~eV from the onset. In contrast, the spectra of two computationally predicted phases are characterized by weak excitations at the onset, growing in intensity with increasing energies, and forming broad absorption bands. The XANES computed from Sb core electrons offers very few signatures for the identification of specific crystal structures and stoichiometries. All Sb K-edge spectra feature broad excitations at the onset, except for \hNaKtwo~and, to a lesser extent, its cubic counterpart, characterized by an intense peak around 1~eV. However, the large broadening associated with such an energetically deep excitation is expected to overshadow these features and to prevent any direct fingerprinting of these compounds. The scenario is even less favorable for the XANES from the Sb L$_2$-edge. In this case, all spectra exhibit broad absorption bands that are hardly distinguishable from each other in a mixed sample. 

Comparing the spectra computed from the BSE and in the IPA reveals the influence of excitonic effects. They are most pronounced in the XANES from Na K-edge, where they give rise to a large redistribution of oscillator strength to the lowest-energy excitations in \Natwo~and at higher energies ($\sim$1~eV) in \NaKtwo. However, in all the considered phases and edges, the differences between the BSE and IPA spectra are so pronounced that it is unlikely that a simplified description of the XANES, neglecting electron-hole correlations, can lead to insightful comparisons with experiments. In all materials, these excitations undergo a red shift larger than 0.5~eV compared to the IPA. In the spectra computed from the potassium K-edge, excitonic effects remain strong but are less relevant than from the Na K-edge. Binding energies are on the order of 200~meV, and the oscillator strength enhancement of low-energy excitations is weakly pronounced except for \hNaKtwo. Stronger excitonic effects are found in the potassium L$_{2,3}$-edge spectra of all considered crystals, especially in the experimental phases \cNatwo~and \hNaKtwo, where they manifest themselves through strong resonances at lowest energies (\cNatwo) and around 1~eV (\hNaKtwo). In the XANES from the Sb K- and L$_2$-edges, excitonic effects are least pronounced: binding energies do not exceed 200~meV, and the oscillator strength enhancement at lowest energies is negligible. These results confirm previous findings~\cite{vorw+17prb,duar-cocc22jpcc}, suggesting a systematic reduction of electron-hole correlation strength with increasing energy of the excited core electrons. This behavior is consistent with physical intuition, whereby transitions from deeper states are subject to a larger screening generated by a larger number of semi-core and valence electrons separating the initial core state of the transition to the final one in the conduction band. It is also worth noting that in the considered MAAs, many-body effects are more pronounced in X-ray spectra than in optical spectra~\cite{amador,xucomppredict}, as predicted by Spicer in 1967~\cite{spic67pr}.

In conclusion, our study provides valuable indications for the X-ray spectroscopic characterization of MAA photocathodes, predicting the spectra not only of experimentally resolved phases but also of computationally predicted polymorphs that could appear as metastable phases in polycrystalline samples.
Our results suggest that excitations from the Na K-edge and the potassium L$_{2,3}$-edge are mostly suited to identify different compositions and crystal structures, while XANES from the potassium K-edge can reveal the presence of hexagonal \NaKtwo, featuring a distinct sharp excitonic resonance. On the contrary, X-ray absorption from Sb K- and L$_2$-edge does not provide any clear fingerprints to identify specific stoichiometries and crystal structures, also given the larger broadening associated with these deep core-level transitions. The absence of experimental references on these compounds prevents a direct comparison between our computational results and measurements, but we are confident that our comprehensive study will stimulate corresponding experiments validating our predictions.

\section*{Acknowledgments}
This work was funded by the German Research Foundation (DFG), Project No. 490940284, DAAD (Program RISE), the German Federal Ministry of Education and Research (Professorinnenprogramm III), and the State of Lower Saxony (Professorinnen f\"ur Niedersachsen). Computational resources were provided by the HPC cluster ROSA at the University of Oldenburg, funded by the DFG (project number INST 184/225-1 FUGG) and by the Ministry of Science and Culture of the Lower Saxony State.

\section*{Data availability statement}
The data that support the findings of this article are openly available at the following Zenodo link: \url{https://doi.org/10.5281/zenodo.15297231}.

\section*{Supporting Information}
In the Supporting Information, the orbital-resolved contributions to the unoccupied density of states of the considered materials are provided (Fig.~S1-S5) along with a table reporting the binding energies of the excitons below the independent-particle onset (Table~S1).

%\bibliography{main}

%apsrev4-2.bst 2019-01-14 (MD) hand-edited version of apsrev4-1.bst
%Control: key (0)
%Control: author (8) initials jnrlst
%Control: editor formatted (1) identically to author
%Control: production of article title (0) allowed
%Control: page (0) single
%Control: year (1) truncated
%Control: production of eprint (0) enabled
\providecommand{\noopsort}[1]{}\providecommand{\singleletter}[1]{#1}%

\end{document}